\documentclass[12pt]{iopart}
\usepackage{graphicx}
\begin{document}

\title{The phonon Hall effect: theory and application}

\author{Lifa~Zhang,$^1$ Jie Ren,$^{2,1}$ Jian-Sheng~Wang,$^{1}$ and Baowen~Li $^{1,2}$}

    \address{$^1$ Department of Physics and Centre for Computational Science and Engineering,
     National University of Singapore, Singapore 117542, Republic of Singapore }
    \address{$^2$ NUS Graduate School for Integrative Sciences and Engineering,
      Singapore 117456, Republic of Singapore}

\date{6 June 2011}
\begin{abstract}
We present a systematic theory of the phonon Hall effect in a ballistic crystal lattice system, and apply it on the kagome lattice which is ubiquitous in various real materials. By proposing a proper second quantization for the non-Hermite Hamiltonian in the polarization-vector space, we obtain a new heat current density operator with two separate contributions: the normal velocity responsible for the longitudinal phonon transport, and the anomalous velocity manifesting itself as the Hall effect of transverse phonon transport. As exemplified in kagome lattices, our theory predicts that the direction of Hall conductivity at low magnetic field can be reversed by tuning temperatures,
which we hope can be verified by experiments in the future. Three phonon-Hall-conductivity singularities induced by phonon-band-topology change are discovered as well, which correspond to the degeneracies at three different symmetric center points, ${\bf \Gamma}$, ${\bf K}$, ${\bf X}$, in the wave-vector space of the kagome lattice.
\end{abstract}

\maketitle

\section{Introduction}
In recent years, phononics, the discipline of science and technology
in processing information by phonons and controlling heat flow,
becomes more and more exciting \cite{Wang2008, wangjs2008}.
Various functional thermal devices such as thermal diode \cite{diode},
thermal transistor \cite{transistor},
thermal logic gates \cite{logic} and thermal memory \cite{memory},
etc., have been proposed to manipulate and control phonons, the
carrier of heat energy and information. And very recently, similar to
the Hall effect of electrons, Strohm \emph{et al.} observed the phonon
 Hall effect (PHE) -- the appearance of a temperature
difference in the direction perpendicular to both the applied
magnetic field and the heat current flowing through an ionic
paramagnetic dielectric sample \cite{strohm05}, which was confirmed
later in Ref.~\cite{inyushkin07}. Such observation of the PHE is really
surprising because phonons as charge-free quasiparticles, different from
electrons, cannot directly couple  to the magnetic field
through the Lorentz force.  Since then,
several theoretical explanations have been proposed
\cite{shengl06,kagan08,wang09,zhang09} to understand this novel phenomenon.
From the work of the PHE in four-terminal nano-junctions and
the phonon Hall conductivity in the two-dimensional
periodic crystal lattice, we know that the PHE can exist even in
the ballistic system.

Geometric phase effects \cite{berry84,xiao10} are fundamentally
important in understanding electrical transport property in
quantum Hall effect \cite{TKNN,kohmoto85},
anomalous Hall effect\cite{AHE,fang03},  and anomalous thermoelectric
transport \cite{xiao06}. It is successful in characterizing the underlying
mechanism of quantum spin Hall effect \cite{shengd06,koenig08}.
Such an elegant connection between mathematics and physics
provides a broad and deep understanding of basic material properties.
Although there is a quite difference between phonons and electrons, we still
can use the topological description to study the underlying
properties of the phonon transport, such as topological phonon
modes in dynamic instability of microtubules \cite{prodan09} and in filamentary structures
\cite{berg11}, Berry-phase-induced
heat pumping \cite{ren10}, and the Berry-phase contribution of
molecular vibrational instability \cite{lv10}.

The topological nature of the PHE is recently studied in Ref.~\cite{zhang10}, where
a general expression for phonon Hall conductivity is obtained in terms of the
Berry curvatures of band structures. In Ref.~\cite{zhang10}, the authors find a phase
transition in the PHE of the honeycomb lattice, explained from
topological nature and dispersion relations. From the Green-Kubo formula and
considering the contributions from all the phonon bands, the authors obtain
the general formula for the phonon Hall conductivity. Then by looking at
the phases of the polarization vectors of both the displacements and
conjugate momenta as a function of the wave vector, a
Berry curvature can be defined uniquely for each band. Combining the above
two steps, at last the phonon Hall conductivity can be written in terms of Berry curvatures. Such derivation gives us a clear picture of
the contribution to the phonon Hall current from all  phonon branches, as well as the
relation between the phonon Hall conductivity and the geometrical phase of the
polarization vectors, which thus helps us to understand the topological picture of the PHE.
However, the process of going from the Berry phase to the heat flux and the
phonon Hall conductivity looks not very clear and natural.

We know that for the Hall effect of the electrons, in addition to the normal
velocity from usual band dispersion contribution, the Berry curvature induces
an anomalous velocity always transverse to the electric field, which gives
rise to a Hall current, thus the Hall effect occurs \cite{xiao10}. For the magnon
Hall effect \cite{katsura10} recently observed, the authors also found the
anomalous velocity due to the Berry connection which is responsible for the
thermal Hall conductivity. Therefore in this article we will derive the theory of
the PHE in a more natural way where the Berry phase effect inducing the anomalous
velocity  contributes to the extra term of the heat current. Thus the Berry phase
effect is straightforward to take the responsibility of the PHE.

A kagome lattice, composed of interlaced triangles whose lattice
points each have four neighboring points \cite{mekata03}, becomes popular in the magnetic
community because the unusual magnetic properties of many real magnetic
materials are associated with those characteristic of the kagome lattice \cite{kagome}.
The schematic figure of kagome lattice is shown in Fig.~\ref{fig1}.
In this paper we also apply the PHE theory to the kagome lattice to investigate
whether the mechanism of the phase transition found in Ref.~\cite{zhang10} is general
and how the phonon Hall conductivity, Chern numbers and the dispersion relation
behave and relate to each other.

In this paper we organize as follows. In Sec.~\ref{sectheory},
we give a new systematic derivation of the theory of the PHE in terms of Berry curvatures.
In this section, we first introduce the Hamiltonian and the modified second quantization,
then derive the heat current density operator which includes both the normal velocity and
the anomalous velocity from the Berry-phase effect. Using the Green-Kubo formula, the
general formula of the phonon Hall conductivity is obtained. Then we give an application
example on the kagome lattice in Sec.~\ref{secapplication}. In this section the computation details
about the dynamic matrix, the Chern numbers and the phonon Hall conductivity are given, and the
behaviors and relations between the phonon Hall conductivity, Chern numbers, and the band structures
are discussed. In the end a short conclusion is presented in Sec.~\ref{seccon}.

\section{The PHE theory}\label{sectheory}
In this section, we give the detailed derivation for the theory of
the PHE. We use the
Hamiltonian in Refs. \cite{zhang10} and \cite{agarwalla11}, which is a positive definite
 Hamiltonian to describe
 the ionic crystal lattice with in an applied magnetic field.
\subsection{The Hamiltonian and the second quantization}
The Hamiltonian for an ionic crystal lattice in a uniform external magnetic field
\cite{zhang10,holz72,agarwalla11} can be written in a compact form as
\begin{eqnarray}
H & = &\frac{1}{2} (p-{ \tilde A }
u)^T (p-{ \tilde A }u) + \frac{1}{2} u^T K u  \qquad\nonumber  \\
& = &\frac{1}{2} p^T p + \frac{1}{2} u^T (K-{ \tilde A }^{2}) u + u^T\!
 { \tilde A }\,p.
\label{eq-model}
\end{eqnarray}
Here, $u$ is a column vector of
displacements from lattice equilibrium positions for all the
degrees of freedom, multiplied by the square root of mass; $p$ is
the conjugate momentum vector, and $K$ is the force constant matrix.
The superscript $T$ stands for the matrix transpose. ${ \tilde A }$ is
an antisymmetric real matrix,  which is block diagonal with elements
 $ \Lambda=
\left(\begin{array}{rr} 0 & h  \\
-h & 0  \\
\end{array}\right)
$ (in two dimensions), where $h$ is proportional to the magnetic field,
and has the dimension of frequency. For simplicity, we will call $h$ magnetic
 field later. According to
\cite{shengl06}, $h$ is estimated to be $0.1\, {\rm cm}^{-1} \approx 3 \times 10^9 \,{ \rm rad\,Hz} $ at a magnetic field $\textbf{\emph{B}} = 1\, {\rm T}$ and a temperature $T = 5.45\,{\rm K}$, which
is within the possible range of the coupling strength in ionic insulators \cite{orbach61,manenkov66}.  The on-site term, $u^T{ \tilde A }p$, can be interpreted as the Raman (or spin-phonon) interaction. Based on quantum theory and symmetry consideration, the phenomenological description of the spin-phonon interaction was proposed many years ago \cite{orbach61,manenkov66,kronig39,vleck40,capellmann87,capellmann89,capellmann91,ioselevich95}. From the first row
of Eq.~(\ref{eq-model}), we find both of the two terms are positive definite,
thus the Hamiltonian ~(\ref{eq-model}) is positive definite.
The origin of the Hamiltonian for the PHE is discussed in
  detail in the supplementary information of Ref.~\cite{zhang10}.

The Hamiltonian Eq.~(\ref{eq-model}) is quadratic in $u$ and $p$.
We can write the linear equation of motion as
\begin{eqnarray}\label{eq-eom1}
\dot p &=&  - (K - \tilde{A}^2 )u - \tilde{A}p, \\
\dot u &=& p - \tilde{A}u.
\label{eq-eom2}
\end{eqnarray}
The equation of motion for the coordinate is,
\begin{equation}
\ddot u + 2\tilde{A}\dot u + K u = 0.
\end{equation}
Since the lattice is periodic, we can apply the Bloch's theorem
$u_l=\epsilon e^{i ({\bf R}_{l} \cdot {\bf k}- \omega t)} $. The
polarization vector $\epsilon$ satisfies
\begin{equation}\label{eq-disper}
\bigl[ (-i\omega + A)^2 + D\bigr] \epsilon = 0,
\end{equation}
where $D({\bf k}) = - A^2+\sum_{l'} K_{ll'} e^{i({\bf R}_{l'} -
{\bf R}_{l})\cdot {\bf k}}$ denotes the dynamic matrix and $A$ is
block diagonal with elements $\Lambda$. $D, K_{l,l'},$ and $A$ are
all $nd\times nd$ matrices, where $n$ is the number of particles
in one unit cell and $d$ is the dimension of the vibration.

From Eq.~(\ref{eq-disper}), we can require the following relations:
\begin{equation}\label{eq-ew1}
\epsilon_{-k}^* = \epsilon_{k}; \;\omega_{-k} = -\omega_{k}.
\end{equation}
Here, we use the short-hand notation $k=({\bf k},\sigma)$ to specify both the
wavevector and the phonon branch, and $-k$ means $(-{\bf k},-\sigma)$. In normal lattice dynamic treatment,
we usually take $\sigma, \omega\geq 0 $ as a convention, and require
$\omega_{\sigma,{\bf k}}=\omega_{\sigma,-{\bf k}}$. For the current problem,
this is not true \cite{zhang10,agarwalla11}. It is more convenient
to have the frequency taking both positive and negative values and require
the above equation (\ref{eq-ew1}).
  And from Eq.~(\ref{eq-eom2}),
the momentum and displacement polarization vectors are related through
\begin{equation}\label{eq-emiu}
\mu_k = -i\omega_k \epsilon_k + A\epsilon_k.
\end{equation}

Equation (\ref{eq-disper}) is not a standard eigenvalue problem.
However, we can describe the system by the polarization vector
 $x =(\mu, \epsilon)^T$, where $\mu $ and $\epsilon $ are associated
  with the momenta and coordinates,
respectively. Using Bloch's theorem, Eqs.~(\ref{eq-eom1}) and
(\ref{eq-eom2}) can be recasted as:
\begin{equation}\label{eq-eom}
i\frac{\partial }
{{\partial t}}x  = H_{\rm eff} x, \;\;\;\;\;\;\;\;\;\;
 H_{\rm eff}= i\left( \begin{array}{cc} -A & -D \\
I_{nd } & -A \end{array} \right).
\end{equation}
Here the $I_{nd}$ is the $nd\times nd$ identity matrix.
Therefore, the eigenvalue problem of the equation of motion
 (\ref{eq-eom}) reads:
\begin{equation}\label{eq-eigen}
H_{\rm eff}\,x_k= \omega_k\,x_k, \;\;\;\;
\tilde{x}_k^T\,H_{\rm eff} = \omega_k\,\tilde{x}_k^T.
\end{equation}
where  the right eigenvector $x_k =(\mu_k,
\epsilon_k)^T$, the left eigenvector ${\tilde x}_k^T =
(\epsilon^\dagger_k, - \mu^\dagger_k)/( - 2i\omega
_k)$, in such choice the second quantization of the Hamiltonian
Eq.~(\ref{eq-model}) holds, which will be proved later. Because the
effective Hamiltonian $H_{\rm eff}$ is not hermitian, the
orthonormal condition then holds between the left and right
eigenvectors,  as
\begin{equation}\label{eq-orthnorm}
{\tilde
x_{\sigma,{\bf k}}}^T\;x_{\sigma',{\bf k}}=\delta_{\sigma\sigma'}.
\end{equation}
We also have the completeness relation as
\begin{equation}\label{eq-complete}
\sum_\sigma x_{\sigma,{\bf k}}\otimes{\tilde
x_{\sigma,{\bf k}}}^T= I_{2nd }.
\end{equation}
The normalization of the eigenmodes is equivalent to \cite{wang09}
\begin{equation}
\epsilon_k^\dagger\,\epsilon_k +
\frac{i}{\omega_k} \epsilon_k^\dagger\, A\,
\epsilon_k  = 1. \label{eq-norm}
\end{equation}

From the eigenvalue problem Eq.~(\ref{eq-eigen}), we know that
the completed set contains the branch of the negative frequency.
And from the topological nature of the PHE \cite{zhang10},
the formula of the phonon Hall conductivity can be written
in the form comprises the contribution of all the branches
including both positive and negative frequency branches.
In order to simplify the notation, for all the branches, we define
\begin{equation}\label{eq-a-k}
a_{-k}=a_{k}^{\dagger}.
\end{equation}
The time dependence of the operators is given by:
\begin{eqnarray}\label{eq-atime1}
a_{k}(t)&=& a_{k} e^{-i \omega_{k}t},\\
a_{k}^{\dagger}(t)&=& a_{k}^{\dagger} e^{i \omega_{k}t}.
\label{eq-atime2}
\end{eqnarray}
The commutation relation is
\begin{equation}\label{eq-acom}
[a_{k},a_{k'}^{\dagger}]= \delta_{k,k'}{\rm sign}(\sigma).
\end{equation}
And we can get
\begin{eqnarray}\label{eq-anum}
\langle a_{k}^{\dagger}a_{k}\rangle&=&f(\omega_{k}){\rm sign}(\sigma);\\
\langle a_{k}a_{k}^{\dagger}\rangle&=&\bigl[1+f(\omega_{k})\bigr]{\rm sign}(\sigma).
\end{eqnarray}
Here $f(\omega _k ) = (e^{\hbar \omega _k/(k_B T)} -
1)^{-1}$ is the Bose distribution function.

The displacement and momentum operators can be written in the
following second quantization forms
\begin{eqnarray}
u_l &=& \sum_k \epsilon_k e^{i {\bf R}_l \cdot {\bf k}}
\sqrt{\frac{\hbar}{2 N |\omega_{k}| }} \, a_k; \\
p_l &=& \sum_k \mu_k e^{i {\bf R}_l \cdot {\bf k}}
\sqrt{\frac{\hbar}{2 N |\omega_{k}| }} \, a_k.
\end{eqnarray}
Here, $|\omega_{k}|=\omega_{k}{\rm sign}(\sigma)$. We can verify that the canonical
commutation relations are satisfied: $[u_l, p_{l'}^T] = i\hbar
\delta_{ll'} I_{nd}$ by using the completeness Eq.~(\ref{eq-complete})
and the commutation relation Eq.~(\ref{eq-acom}).
The Hamiltonian Eq.~(\ref{eq-model}) then can be written as \cite{agarwalla11}
\begin{equation}
H=\frac{1}{2}\sum_{l,l'} \tilde{\chi}^T_{l} \left( \begin{array}{cc}
A \delta_{l,l'} & K_{l,l'}-A^2\delta_{l,l'} \\
-I_{nd} \delta_{l,l'} & A \delta_{l,l'} \end{array} \right) \chi_{l'}
\end{equation}
where
\begin{eqnarray} \label{eq-xl}
\chi_{l}=\left( \begin{array}{rr}p_{l} \\ u_{l} \end{array} \right)
&=&\sqrt{\frac{\hbar}{N}}\sum_k x_k e^{i {\bf R}_l \cdot {\bf k}} c_k\; a_k; \\
\tilde{\chi}_{l} = \left( \begin{array}{rr}u_{l} \\ -p_{l} \end{array} \right)
&=&\sqrt{\frac{\hbar}{N}}\sum_k \tilde x_k e^{-i {\bf R}_l \cdot {\bf k}}
\tilde{c}_k\; a_k^{\dagger}.\label{eq-txl}
\end{eqnarray}
 Here $c_k=\sqrt{\frac{1}{2 |\omega_{k}|}}$ and
$\tilde{c}_k=( - 2i\omega _k )\sqrt{\frac{1}{2|\omega_{k}| }}$.
It is easy to verify that $[{\chi}_{l},\tilde{\chi}^T_{l'}]=-i\hbar
\delta_{ll'}I_{2nd }$.

Because of $e^{i ({\bf R}_{l'} \cdot {\bf k'}-{\bf R}_l \cdot {\bf k})}
=e^{i ({\bf R}_l \cdot ({\bf k'}-{\bf k})+({\bf R}_{l'}- {\bf R}_l)
\cdot {\bf k'})}$ and the definition of the dynamic matrix $D$,
then the Hamiltonian can be written as
\begin{eqnarray}\label{eq-ham2q}
H&=&\frac{\hbar}{2N}\sum_{k,k',l}e^{i {\bf R}_l \cdot
({\bf k'}-{\bf k})} \tilde c_k\,c_{k'} \tilde x_k^T
\left( \begin{array}{cc} A  & D({\bf k'}) \\  -I_{nd} & A \end{array} \right)
x_{k'} a_k^\dagger  a_{k'} \nonumber\\
&=&\frac{\hbar}{2N}\sum_{k,k',l} e^{i {\bf R}_l \cdot ({\bf k'}-{\bf k})}
\tilde c_k\,c_{k'} \tilde x_k^T i H_{\rm eff}x_{k'} a_k^\dagger  a_{k'}\nonumber\\
&=&\frac{1}{2}\sum_k \hbar |\omega_{k}|a_k^\dagger  a_k,
\end{eqnarray}
which contains both the positive and negative branches.
Here we use the identity $\sum_{l}e^{i {\bf R}_l \cdot ({\bf k'}-{\bf k})}
=N\delta_{\bf k'k}$ and the eigenvalue problem Eq.~(\ref{eq-eigen}).
Using the relations Eqs.~(\ref{eq-a-k}) and (\ref{eq-acom}), it is easy
to prove that
Eq. (\ref{eq-ham2q}) is equivalent to the form $H=\sum_{\sigma>0,{\bf k}} \hbar
\omega_k (a_k^\dagger a_k +
1/2)$ which only includes the nonnegative branches.
\subsection{The heat current operator}
The heat current density can be computed as \cite{hardy63}:
\begin{equation}
{\bf J}= \frac{1}{2V} \sum_{l,l'} ({\bf R}_l\! -\! {\bf R}_{l'})
u^T_{l} K_{ll'} \dot{u}_{l'},
\end{equation}
where $V$ is the total volume of $N$ unit cells. Because of the
equation of motion Eq.~(\ref{eq-eom2}), we can rewrite the heat current as
\begin{equation}
{\bf J}= \frac{1}{4V} \sum_{l,l'} \tilde{\chi}^T_{l} {\bf M}_{l\,l'} \chi_{l'},
\end{equation}
where
\begin{equation}
 {\bf M}_{l\,l'} =\left( \begin{array}{cc} ({\bf R}_{l}-
 {\bf R}_{l'})K_{ll'} & -({\bf R}_{l}- {\bf R}_{l'})(K_{ll'}A+AK_{ll'}) \\
 0 &  ({\bf R}_{l}- {\bf R}_{l'})K_{ll'}\end{array} \right)
\end{equation}
Inserting the Eqs.~(\ref{eq-xl},\ref{eq-txl}), we obtain
\begin{equation}
{\bf J}= \frac{\hbar}{4VN} \sum_{k,k',l,l'}\tilde{c}_k c_{k'}
e^{i ({\bf R}_{l'} \cdot {\bf k'}-{\bf R}_l \cdot {\bf k})}
\tilde{x}^T_{k} {\bf M}_{l\,l'} x_{k'} a_k^\dagger  a_{k'},
\end{equation}
Because of
\begin{equation}
\sum_{l}e^{i {\bf R}_l \cdot ({\bf k'}-{\bf k})}\sum_{l'}
e^{i({\bf R}_{l'}- {\bf R}_l)\cdot {\bf k'}}({\bf R}_{l}- {\bf R}_{l'})K_{ll'}
=i N\delta_{\bf k'k}\frac{\partial D}{\partial {\bf k'}},
\end{equation}
the heat current can be written as
\begin{equation}
{\bf J}= \frac{i\hbar}{4V} \sum_{\sigma,\sigma',{\bf k}}
\tilde{c}_{\sigma,{\bf k}} c_{\sigma',{\bf k}} \tilde{x}^T_{\sigma,{\bf k}}
\frac{\partial H_{\rm eff}^2}{\partial {\bf k}}
  x_{\sigma',{\bf k}} a_{\sigma,{\bf k}}^\dagger  a_{\sigma',{\bf k}},
\end{equation}
here we use
\begin{equation}\label{eq-dh2dk}
\frac{\partial H_{\rm eff}^2}{\partial {\bf k}}=\left( \begin{array}{cc}
\frac{\partial D}{\partial {\bf k}} & -(A\frac{\partial D}{\partial {\bf k}}
+\frac{\partial D}{\partial {\bf k}}A) \\
0 &  \frac{\partial D}{\partial {\bf k}}\end{array} \right)
\end{equation}
by making the first derivative of the square of the effective
Hamiltonian Eq.~(\ref{eq-eom}) with respect to the wave vector ${\bf k}$.
From the eigenvalue problem Eq.~(\ref{eq-eigen}), we have
\begin{equation}
H_{\rm eff} X = X \Omega;\;\; \tilde{X}^T H_{\rm eff}  = \Omega \tilde{X}^T.
\end{equation}
Where the ${2nd}\times {2nd}$ matrices $X=(x_1,x_2,...,x_{2nd})=\{x_{\sigma}\}$ (the system has
 ${2nd}$ phonon branches), $\tilde{X}=\{\tilde{x}_{\sigma}\}$,
and $\Omega={\rm diag}(\omega_1,\omega_2,...,\omega_{ 2nd})=\{\omega_{\sigma}\}$.
Because of the completeness relation Eq.~(\ref{eq-complete}), $X \tilde{X}^T= I_{2 nd}$, we get
\begin{equation}\label{eq-hamsqr}
H_{\rm eff}^2 = X \Omega^2 \tilde{X}^T.
\end{equation}
By calculating the derivative of the above equation, and using the
definition of Berry connection,
\begin{equation} \label{eq-brycnct}
{\bf \mathcal{A}}=\tilde{X}^T \frac{\partial X}{\partial {\bf k}}.
\end{equation}
Taking the first derivative of  Eq.~(\ref{eq-hamsqr}) with respect to ${\bf k}$,
we obtain
\begin{equation}
\frac{\partial H_{\rm eff}^2}{\partial {\bf k}}
=X\left(\frac{\partial \Omega^2}{\partial {\bf k}}
+[{\bf \mathcal{A}}, \Omega^2]\right)\tilde{X}^T.
\end{equation}
Because of the orthogonality relation between left and right eigenvector
Eq.~(\ref{eq-orthnorm}),  at last we obtain the heat current as
  \begin{equation}\label{eq-current}
{\bf J}= \frac{i\hbar}{4V} \sum_{\sigma,\sigma',{\bf k}}\tilde{c}_{\sigma,{\bf k}}
 c_{\sigma',{\bf k}} a_{\sigma,{\bf k}}^\dagger
 \left(\frac{\partial \Omega^2}{\partial {\bf k}}+[{\bf \mathcal{A}},
 \Omega^2]\right)_{\sigma,\sigma'} a_{\sigma',{\bf k}}.
\end{equation}
The first term $\frac{\partial \Omega^2}{\partial {\bf k}}$ in the bracket
is a diagonal one corresponding to $\omega_\sigma \frac{\partial \omega_\sigma}
{ \partial {\bf k}}$ relating the group velocity. The second term in the
bracket $[{\bf \mathcal{A}}, \Omega^2]$ gives the off-diagonal elements of
the heat current density,  which can be regarded as the contribution from
anomalous velocities similar to the one in the intrinsic anomalous Hall effect.
 The Berry connection ${\bf \mathcal{A}}$, or we can call it Berry vector
 potential matrix (the Berry vector potential defined in Ref.~\cite{zhang10},
 ${\bf A}^\sigma ({\bf k})$, is equal to $i {\bf \mathcal{A}}^{\sigma\sigma}=i
 \tilde{x}_\sigma^T \frac{\partial x_\sigma}{\partial {\bf k}}$), induces
 the anomalous velocities to the heat current, which will take the responsibility
 of the PHE.  Therefore, the Berry vector potential comes naturally into
 the heat current and the PHE. Such a picture is clearer than that in Ref.~\cite{zhang10}.
 \subsection{The phonon Hall conductivity}
Inserting the coefficients $\tilde{c}$ and $c$ to Eq.~(\ref{eq-current}), we get
\begin{equation}
{\bf J}= \frac{\hbar}{4V} \sum_{\sigma,\sigma',{\bf k}}
\frac{\omega_{\sigma,{\bf k}}}{\sqrt{|\omega_{\sigma,{\bf k}}
\omega_{\sigma',{\bf k}}|}} a_{\sigma,{\bf k}}^\dagger
\left(\frac{\partial \Omega^2}{\partial {\bf k}}+[{\bf \mathcal{A}},
\Omega^2]\right)_{\sigma,\sigma'} a_{\sigma',{\bf k}}.
\end{equation}
This expression is equivalent to that given in Refs.~\cite{zhang10} and \cite{agarwalla11}.
Based on such expression of heat current, the phonon Hall
conductivity can be obtained through the Green-Kubo formula
\cite{mahan00}:
\begin{equation}
\kappa_{xy} = \frac{V}{\hbar T} \int_0^{\hbar/(k_B T)}\!\!\!\! d\lambda
\int_0^\infty\! dt\, \bigl\langle J^x(-i\lambda) J^y(t) \bigr\rangle_{\rm eq},
\label{eq-GK}
\end{equation}
where the average is taken over the equilibrium ensemble with Hamiltonian $H$.
The time dependence of the creation and annihilation operators are given as
 Eqs.~(\ref{eq-atime1}) and (\ref{eq-atime2}),
which are also true if $t$ is imaginary. From the Wick theorem, we have
\begin{eqnarray}
\langle a_{\sigma,{\bf k}}^\dagger a_{\sigma',{\bf k}}
a_{\bar{\sigma},{\bf \bar{k}}}^\dagger a_{\bar{\sigma'},{\bf \bar{k}}}\rangle
&=& \langle a_{\sigma,{\bf k}}^\dagger a_{\sigma',{\bf k}} \rangle
\langle a_{\bar{\sigma},{\bf \bar{k}}}^\dagger
a_{\bar{\sigma'},{\bf \bar{k}}}\rangle \nonumber \\
 &+&\langle a_{\sigma,{\bf k}}^\dagger
 a_{\bar{\sigma},{\bf \bar{k}}}^\dagger \rangle
 \langle a_{\sigma',{\bf k}}  a_{\bar{\sigma'},{\bf \bar{k}}}\rangle \nonumber \\
 &+&\langle a_{\sigma,{\bf k}}^\dagger
 a_{\bar{\sigma'},{\bf \bar{k}}} \rangle
 \langle a_{\sigma',{\bf k}} a_{\bar{\sigma},{\bf \bar{k}}}^\dagger \rangle.
\end{eqnarray}
Using the properties of the operators $a^\dagger$ and $a$ as Eq.~(\ref{eq-anum}),
we have
\begin{equation}
\begin{array}{ll}
\langle a_{\sigma,{\bf k}}^\dagger a_{\sigma',{\bf k}}
\rangle \langle a_{\bar{\sigma},{\bf \bar{k}}}^\dagger
a_{\bar{\sigma'},{\bf \bar{k}}}\rangle &   \\
= f(\omega_{\sigma,{\bf k}})
f(\omega_{\bar{\sigma},{\bf \bar{k}}})\delta_{\sigma \sigma'}
\delta_{\bar{\sigma} \bar{\sigma'}}{\rm sign}(\sigma)
{\rm sign}(\bar{\sigma}), &\\
\langle a_{\sigma,{\bf k}}^\dagger
a_{\bar{\sigma},{\bf \bar{k}}}^\dagger \rangle
 \langle a_{\sigma',{\bf k}}  a_{\bar{\sigma'},{\bf \bar{k}}}\rangle &\\
 = f(\omega_{\sigma,{\bf k}})(f(\omega_{\sigma',{\bf k}})+1)
 \delta_{\bf\bar{k},-k}\delta_{\sigma,-\bar{\sigma}}
 \delta_{\sigma',-\bar{\sigma'}}{\rm sign}(\sigma){\rm sign}(\sigma') &\\
\langle a_{\sigma,{\bf k}}^\dagger
 a_{\bar{\sigma'},{\bf \bar{k}}} \rangle
 \langle a_{\sigma',{\bf k}} a_{\bar{\sigma},{\bf \bar{k}}}^\dagger \rangle &\\
  = f(\omega_{\sigma,{\bf k}})(f(\omega_{\sigma',{\bf k}})+1)
  \delta_{\bf\bar{k},k}\delta_{\sigma,\bar{\sigma'}}
  \delta_{\sigma',\bar{\sigma}}{\rm sign}(\sigma){\rm sign}(\sigma'). &
  \end{array}
\end{equation}

Similar as that in Ref.~\cite{zhang10}, the diagonal term  $\frac{\partial \Omega^2}{\partial {\bf k}}$ in the bracket
corresponding to $\omega_\sigma \frac{\partial \omega_\sigma}
{ \partial {\bf k}}$ has no contribution to the phonon Hall conductivity because which
is an odd function of ${\bf k}$. Because of the off-diagonal term
\begin{equation}
[\mathcal{A}_{k_\alpha},
\Omega^2]_{\sigma,\sigma'}=(\omega_{\sigma'}^2-\omega_{\sigma}^2)
\mathcal{A}_{k_\alpha}^{\sigma\sigma'}
\end{equation}
and $\mathcal{A}_{k_\alpha}^{\sigma\sigma'}=
\tilde{x}_\sigma^T \frac{\partial x_{\sigma'}}{\partial {k_\alpha}}$ from the definition,
the phonon Hall conductivity can be written as
\begin{eqnarray}
\kappa_{xy}& = &
\frac{\hbar }
{{8V  T}}\sum_{{\bf k},\sigma,\sigma'\neq\sigma} [f(\omega _\sigma )
- f(\omega _{\sigma '})](\omega _\sigma  + \omega _{\sigma '})^2 \nonumber\\
&  & \times \frac{i} {4 \omega _\sigma  \omega _{\sigma '} }
\frac{{\epsilon _\sigma ^\dag  \frac{{\partial D}}
{{\partial k_x }}\epsilon _{\sigma '} \epsilon _{\sigma '}^\dag
\frac{{\partial D}}
{{\partial k_y }}\epsilon _\sigma  }}
{{(\omega _\sigma   - \omega _{\sigma '} )^2 }}.
\end{eqnarray}
Here we simplify the notation of the subscripts of $\omega, \epsilon$
which have the same wave vector ${\bf k}$.  We can prove
$\kappa_{xy}=-\kappa_{yx}$, such that
\begin{equation}\label{eq-kxy}
\kappa_{xy} =
\frac{\hbar }
{{16V  T}}\sum_{{\bf k},\sigma,\sigma'\neq\sigma} {[f(\omega _\sigma )
- f(\omega _{\sigma '})](\omega _\sigma  + \omega _{\sigma '})^2
B_{k_x k_y}^{\sigma \sigma '}},
\end{equation}
here
\begin{eqnarray}
B_{k_x k_y}^{\sigma \sigma '}&=&\frac{i}
{{4 \omega _\sigma  \omega _{\sigma '} }}\frac{\epsilon_\sigma^\dag\frac{{\partial D}}
{{\partial k_x }}\epsilon _{\sigma '} \epsilon _{\sigma '}^\dag
\frac{{\partial D}}
{{\partial k_y }}\epsilon _\sigma - (k_x  \leftrightarrow k_y ) }
{(\omega _\sigma   - \omega _{\sigma '} )^2 }. \nonumber\\
&=&i\frac{\tilde{x}_\sigma^T\frac{{\partial H_{\rm eff}}}
{{\partial k_x }}x_{\sigma '} \tilde{x}_{\sigma'}^T
\frac{{\partial H_{\rm eff}}}
{{\partial k_y }}x_\sigma - (k_x  \leftrightarrow k_y ) }
{(\omega _\sigma   - \omega _{\sigma '} )^2 }. \nonumber\\
&=&-i\left(\mathcal{A}_{k_x}^{\sigma \sigma '}
\mathcal{A}_{k_y}^{\sigma' \sigma}-(k_x  \leftrightarrow k_y )\right),
\label{eq-curv}
\end{eqnarray}
in the last step we use the relation
$\tilde{x}_\sigma^T\frac{{\partial H_{\rm eff}}}
{{\partial k_x }}x_{\sigma '}=(\omega _{\sigma'}   - \omega _{\sigma} )
\tilde{x}_\sigma^T\frac{{\partial }}
{{\partial k_x }}x_{\sigma '}$ and the definition of ${\bf \mathcal{A}}$
in Eq.~(\ref{eq-brycnct}).
And the Berry curvature is
\begin{eqnarray}
B _{k_x k_y }^\sigma
 &= &\sum_{\sigma ' \ne \sigma } B _{k_x k_y }^{\sigma \sigma '} \nonumber\\
 &= & -i\sum_{\sigma ' }\left( \mathcal{A}_{k_x}^{\sigma \sigma '}
 \mathcal{A}_{k_y}^{\sigma' \sigma}-(k_x  \leftrightarrow k_y )\right) \nonumber\\
 &= & i \left( \frac{\partial}{\partial k_x}\mathcal{A}_{k_y}^{\sigma \sigma }
 -(k_x  \leftrightarrow k_y )\right)
\end{eqnarray}
The definition of Berry curvature here is the same as that of Ref.~\cite{zhang10}, that is,
$B _{k_x k_y }^\sigma=\frac{\partial }
{{\partial k_x }}{\bf A}_{k_y }^\sigma   - \frac{\partial }
{{\partial k_y }}{\bf A}_{k_x }^\sigma$.
From the above derivation, we find that a Berry curvature can be defined
uniquely for each band by looking at the phases of the polarized vectors
 of both the displacements and conjugate momenta as functions of the wave vector.
If we only look at the polarized vector $\epsilon$ of the displacement, a
Berry curvature cannot properly be defined. We need both $\epsilon$ and $\mu$.
The nontrivial Berry vector potential takes the responsibility of the PHE.
The associated topological Chern number is obtained through integrating
the Berry curvature over the first Brillouin zone as
\begin{equation}\label{eq-cherni}
C^\sigma   = \frac{1} {{2\pi }}\int_{{\rm BZ}} {dk_x dk_y B
_{k_x k_y }^\sigma  }= \frac{{2\pi }} {{L^2}}\sum\limits_{\bf k}
{B _{k_x k_y }^\sigma  },
\end{equation}
where, $L$ is the length of the sample.

\section{Application on the kagome lattice}
\label{secapplication}
\begin{figure}
\centerline{\includegraphics[width=0.6\columnwidth]{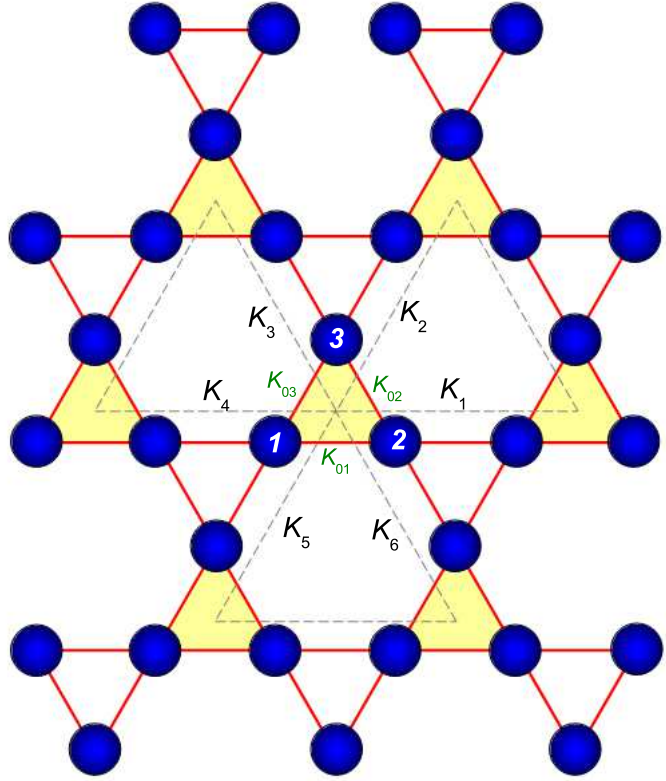}}
\caption{ \label{fig1} (color online) The schematic picture of kagome
lattice. Each unit cell has three atoms such as the number shown 1,2,3.
The coupling between
the atoms are $K_{01}, K_{02}, K_{03}$. Each unit cell has six nearest
neighbors; the coupling between the unit cell and the neighbors are
$K_1, K_2, ..., K_6$.  }
\end{figure}
In Ref.~\cite{zhang10}, we provide a topological understanding of
the PHE in dielectrics with Raman spin-phonon coupling for the honeycomb
lattice structure. Because of  the nature of phonons, the phonon Hall
conductivity, which is not directly proportional to the Chern number,
is not quantized. We observed a phase transition in the PHE, which
corresponds to the sudden change of band topology, characterized by
the altering of integer Chern numbers. Such PHE can be explained by
touching and splitting of phonon bands. To check whether the mechanism
of the PHE is universal, in the following we apply the theory to the
kagome lattice, which has been used to model many real materials
\cite{kagome}.

\subsection{Calculation of the dynamic matrix $D$}
In order to calculate the phonon Hall conductivity, we first need to calculate the dynamic
matrix $D({\bf k})$, for
the two-dimensional kagome lattice. As shown in Fig.~\ref{fig1},
each unit cell has three atoms, thus $n=3$. We only consider
the nearest neighbor interaction. The spring constant matrix
along $x$ direction is assumed as
\begin{equation}
K_x=\left(\begin{array}{cc} K_L & 0  \\
0 & K_T  \\
\end{array}\right).
\end{equation}
$K_L = 0.144\,$eV/(u\AA$^2$) is the longitudinal spring constant and
the transverse one $K_T$ is 4 times smaller.  The unit cell lattice
vectors are $(a,0)$ and $(a/2, a\sqrt{3}/2)$ with $a=1\,$\AA.

To obtain the explicit formula for the dynamic matrix, we first
define a rotation operator in two dimensions as:
$$
U(\theta ) = \left( {\begin{array}{*{20}c}
   {\cos \theta } & { - \sin \theta }  \\
   {\sin \theta } & {\cos \theta }  \\
 \end{array} } \right).
$$
The three kinds of spring-constant matrices between two atoms are
$K_{01}  = K_x $ (between atoms 1 and 2 in Fig.~\ref{fig1}),
$K_{02}  = U(\pi /3)K_x U( -\pi /3)$ (between atoms 2 and 3),
$K_{03}  = U( - \pi /3)K_x U(\pi /3)$ (between atoms 3 and 1), which are
$2\times2$ matrices. Then we can obtain the on-site spring-constant
matrix and the six spring-constant matrices between the unit cell
and its nearest neighbors as:
$$
  K_0  = \left( {\begin{array}{ccc}
   2(K_{01}  + K_{02}) & - K_{01} & - K_{02}  \\
   - K_{01}  &2(K_{01}  + K_{03})& - K_{03} \\
   - K_{02}  & - K_{03} & 2(K_{02}  + K_{03})
 \end{array} } \right),$$
 \begin{eqnarray}
  K_1 & = &\left( {\begin{array}{ccc}
   0 & 0 & 0  \\
   - K_{01}  &0& 0\\
   0 &0 & 0
 \end{array} } \right),\;
  K_2  = \left( {\begin{array}{ccc}
   0 & 0 & 0 \\
   0 & 0 & 0 \\
   - K_{02}  & 0 & 0
 \end{array} } \right),\nonumber  \\
  K_3  & =& \left( {\begin{array}{ccc}
   0 & 0 & 0  \\
   0 &0& 0\\
   0& - K_{03} & 0
 \end{array} } \right),\;
   K_4 = \left( {\begin{array}{ccc}
   0 & - K_{01} & 0 \\
   0 &0& 0\\
   0& 0 & 0
 \end{array} } \right), \nonumber  \\
   K_5 &=&\left( {\begin{array}{ccc}
   0 & 0 & - K_{02}  \\
   0 &0& 0\\
   0 & 0& 0
 \end{array} } \right),\;
   K_6= \left( {\begin{array}{ccc}
   0 & 0 & 0  \\
   0&0& - K_{03} \\
   0 & 0 & 0
 \end{array} } \right),\nonumber
 \end{eqnarray}
which are $6\times6$ matrices.
Finally we can obtain the $6\times6$ dynamic matrix $D({\bf k})$ as
\begin{eqnarray}\label{eq-Dexplict}
D({\bf k}) & =&-A^2+ K_0  + K_1 e^{ik_x }  + K_2 e^{i(\frac{k_x}{2} +
\frac{\sqrt3 k_y } {2})}\nonumber \\  & &+ K_3 e^{i(-\frac{k_x}{2} +
\frac{\sqrt3 k_y } {2})} + K_4 e^{-ik_x }  \nonumber \\
& & + K_5 e^{i(-\frac{k_x}{2} - \frac{\sqrt3 k_y } {2})}
+ K_6 e^{i(\frac{k_x}{2} - \frac{\sqrt3 k_y } {2})},
\end{eqnarray}
where, $A^2=-h^2\cdot I_6$, here $I_6$ is the $6\times6$ identity
matrix.

\begin{figure}[t]
\centerline{\includegraphics[width=0.9\columnwidth]{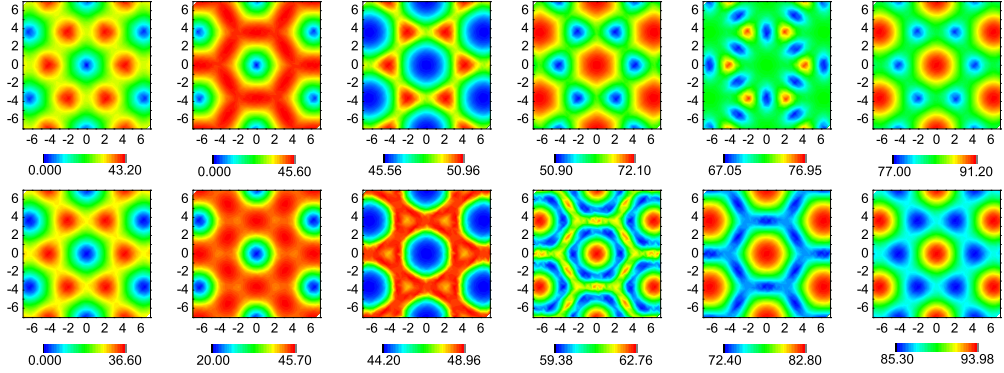}}
\caption{ \label{fig2} (color online) The contour map of dispersion
relations for the positive frequency bands. For all the insets, the
horizontal and vertical axes correspond to wave vector $k_x$ and $k_y$,
respectively. The upper six insets are the dispersion relations for bands
1 to 6 (from left to right) at $h=0$, respectively. And $h=10$ rad/ps for
the lower ones.  }
\end{figure}

\subsection{The PHE and the associated phase transition }
After we get the expression for the dynamic matrix, we can calculate
the eigenvalues and eigenvectors of the effective Hamiltonian. Inserting
the eigenvalues, eigenvectors and the $D$ matrix to the formula
Eq.~(\ref{eq-kxy}), we are able to compute the phonon Hall conductivity.
As is well known, in quantum Hall effect for
electrons, the Hall conductivity is just the Chern number in units
of $e^2/h$ ($h$ is the Planck constant); thus with the varying of
magnetic field, the abrupt change of Chern numbers directly induces
the obvious discontinuity of the Hall conductivity. However, for the
PHE, there is an extra weight of $(\omega _\sigma  +
\omega _{\sigma '})^2$ in Eq.~(\ref{eq-kxy}), which can not be moved out from the
summation. As a consequence, the change of phonon Hall conductivity
is smoothened at the critical magnetic field. However, in the study on the
PHE in the honeycomb lattice system \cite{zhang10}, from the first
derivative of phonon Hall conductivity with respect to the magnetic
field $h$, at the critical point $h_c$, we still can observe the
divergence (singularity) of $d\kappa_{xy}/dh$, where the phase
transition occurs corresponding to the sudden change of the Chern
numbers. Can such mechanism be applied for the kagome lattice system?
In the following, we give a detailed discussion on it.

Inserting the dynamic matrix Eq.~(\ref{eq-Dexplict}) to the effective Hamiltonian
Eq.~(\ref{eq-eigen}), we calculate eigenvalues and eigenvectors of the system, and
also get the dispersion relation of the system. Because each unit cell has
three atoms, and we only consider the two-dimensional motion, we get
six phonon branches with positive frequencies. The branches with negative frequencies
have similar behavior because of $\omega_{-k}=-\omega_{k}$. We show the contour
map of the dispersion relation in Fig.~\ref{fig2}. We can see that the dispersion
relations have a 6-fold symmetry. For different bands, they are different.
With a changing magnetic field, the dispersion relations vary. The point
 ${\bf \Gamma}$ (${\bf k}=(0,0)$) is the 6-fold symmetric center; the point
  ${\bf K}$ (${\bf k}=(\frac{4\pi}{3},0)$) is 3-fold symmetric center; and the
   middle point of the line between two 6-fold symmetric centers, ${\bf X}$
   (${\bf  k}=(\pi,\frac{\sqrt{3}\pi}{3})$) is a 2-fold symmetric center.
   In the following discussion, we will see the possible bands touching at
    these symmetric centers.

\begin{figure}[t]
\centerline{\includegraphics[width=0.9\columnwidth]{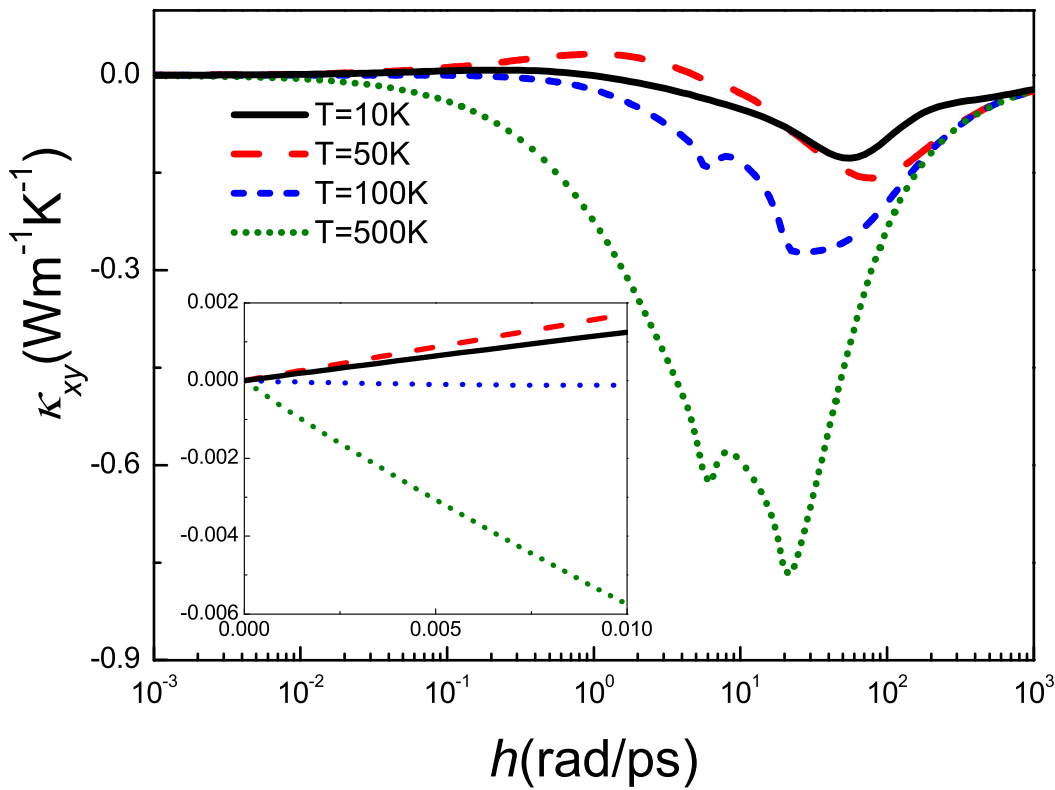}}
\caption{ \label{fig3} (color online) The phonon Hall conductivity
 vs magnetic field at different temperatures. The inset is the zoom-in
 curve of the phonon Hall conductivity at weak magnetic field.
 Here the sample size $N_L$=400. }
\end{figure}

Using the formula  Eq.~(\ref{eq-kxy}), we calculate the phonon Hall conductivity
 of the kagome lattice systems, the results are shown in Fig.~\ref{fig3}. Similar as shown in
 Ref.~\cite{zhang10}, we find the nontrivial behavior of the phonon Hall
 conductivity as a function of the magnetic field. When $h$ is small,
 $\kappa_{xy}$ is proportional to $h$, which is shown in the inset of
 Fig.\ref{fig3}; while the dependence becomes nonlinear when  $h$ is large.
 As $h$ is further increased, the magnitude of $\kappa_{xy}$ increases before
it reaches a maximum magnitude at  certain value of $h$.
Then the magnitude of $\kappa_{xy}$ decreases and goes to zero at very large $h$.
  The on-site term $\tilde{A}^2$ in the Hamiltonian (\ref{eq-model}) increases
  with $h$ quadratically so as to blockade the phonon transport, which competes
  with the spin-phonon interaction. Because of the coefficient of
   $f(\omega _\sigma )$ in the summation of the formula Eq.~(\ref{eq-kxy}),
   the sign of the Hall conductivity will change with temperatures, which is clearly shown in the inset of Fig.\ref{fig3}. While the phonon hall conductivity at weak magnetic field is always positive for the honeycomb lattice, the sign reverse of the phonon Hall conductivity with temperature for the kagome lattices is novel and interesting, which could be verified by future experimental measurements.

\begin{figure}[t]
\centerline{\includegraphics[width=0.9\columnwidth]{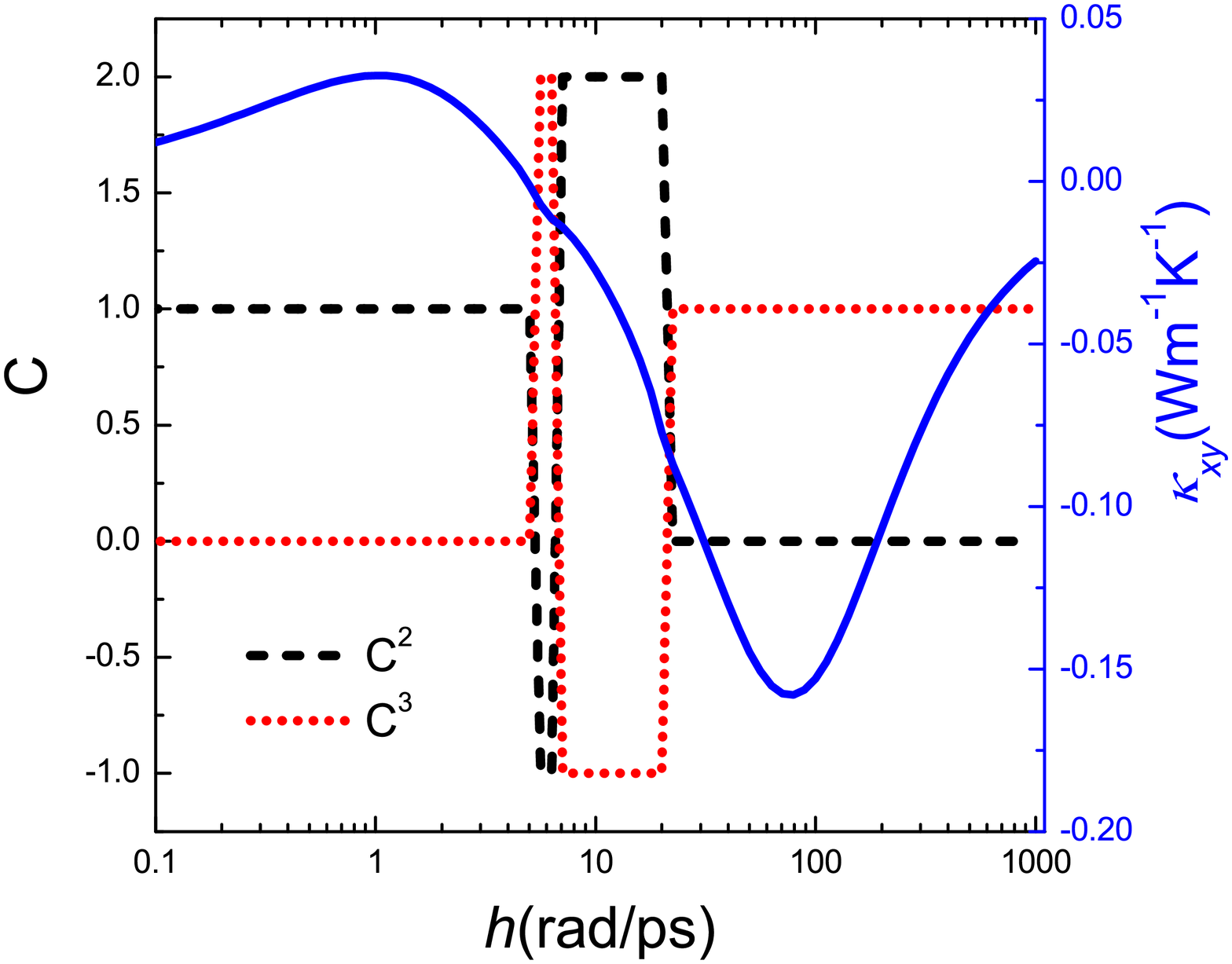}}
\caption{ \label{fig4} (color online) The Chern numbers and the phonon Hall
conductivity vs magnetic field. The dashed line and the dotted line correspond
to the Chern numbers of phonon bands 2 and 3 (left scale). The solid line
correspond to the phonon Hall conductivity (right scale) at $T=50$ K. }
\end{figure}
We plot the curves of the Chern numbers of bands 2 and 3 as a function of
the magnetic field in Fig.~\ref{fig4}. The phonon Hall conductivity at
$T=50 K$ is also shown for comparison. To calculate the integer Chern numbers,
large number of ${\bf k}$-sampling points $N$ is needed. However there is always a zero
eigenvalue at the ${\bf \Gamma}$  point of the dispersion relation, which corresponds to
a singularity of the Berry curvature. Therefore, we cannot sum up
the Berry curvature very near this point to obtain Chern number of
this band, unless we add a negligible on-site potential
$\frac{1}{2}u^TV_{\mathrm{onsite}}u$ to the original Hamiltonian
\cite{zhang10}, which will not change the topology of the space of
the eigenvectors. In Fig.~\ref{fig4}, we set
$V_{\mathrm{onsite}}=10^{-3} K_L$.   The Chern numbers of
bands 2 and 3 have three jumps with the increasing of the magnetic
field, although the phonon Hall conductivity is continuous.
For other bands, the Chern numbers keep constant: $C^1=C^4=-1$, $C^5=0$,
and $C^6=1$. For the electronic Hall effect, we know it is quantized because
the Hall conductivity is directly proportional to the quantized Chern numbers.
Here we also find the quantized effect of the Chern numbers from
Fig.~\ref{fig3}, while there is no quantized effect for the phonon
Hall conductivity. Such difference of the PHE from the electronic Hall
effect comes from the different nature of the phonons respective to the
electrons. In Eq.~(\ref{eq-kxy}), in the summation, an extra term
$ (\omega _\sigma  + \omega _{\sigma '})^2 $  relating to the phonon
energy which is an analog of the electrical charge term $e^2$ in the
electron Hall effect, can not be moved out from the summation. Combining
the Bose distribution, the term $f(\omega _\sigma ) (\omega _\sigma  +
\omega _{\sigma '})^2 $ make the phonon Hall conductivity smooth,
no discontinuity comes out although the Chern numbers have some
sudden jumps. From the discussion in Ref.~\cite{zhang10},
the discontinuity of the Chern numbers corresponds to the phase transitions
and would relate to the divergency of derivative of the phonon Hall conductivity.

\begin{figure}
\centerline{\includegraphics[width=0.9\columnwidth]{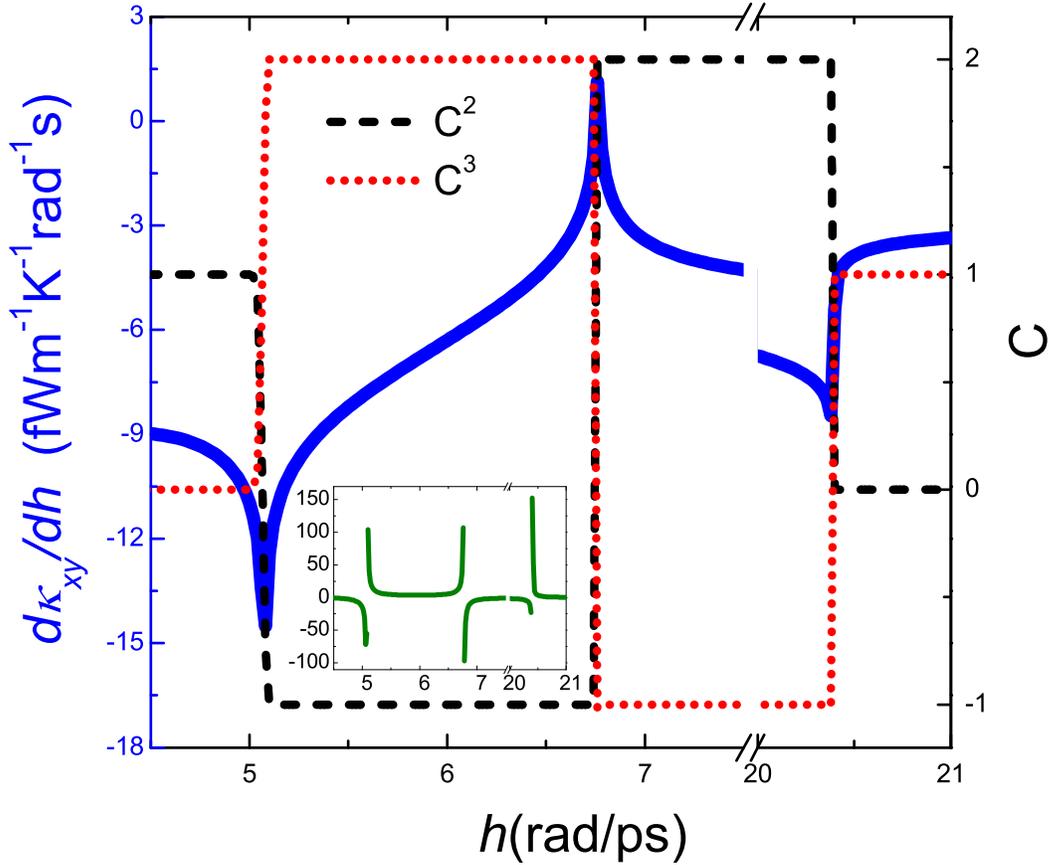}}
\caption{ \label{fig5} (color online) The first derivative of the phonon
Hall conductivity $dk_{xy}/dh$ at $T=50 K$ and the Chern numbers of bands
2 and 3 in the vicinity of the magnetic fields. The  solid line correspond
to the $dk_{xy}/dh$ at $T=50$K (left scale); the dashed and dotted lines
correspond to the Chern numbers of bands 2 and 3, respectively (right scale).
The inset shows the second derivative with respective to the magnetic field
$dk_{xy}^2/dh^2$ (vertical axis) vs magnetic field $h$ (horizontal axis)
 at $T=50$ K.  }
\end{figure}

\begin{figure}
\centerline{\includegraphics[width=0.9\columnwidth]{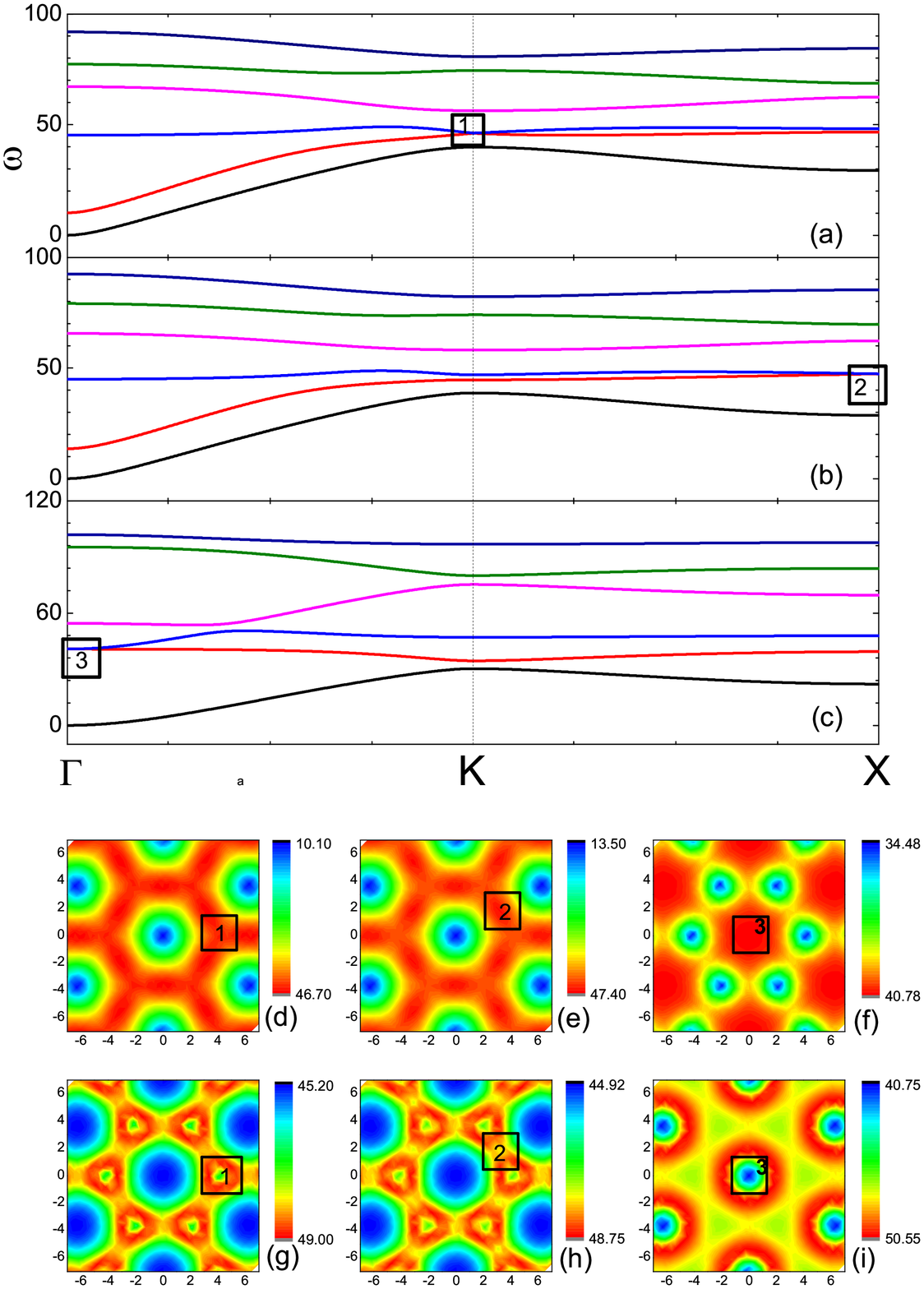}}
\caption{ \label{fig6} (color online) The dispersion relations around
the critical magnetic fields. (a), (b), and (c) show the dispersion
relations along the direction from ${\bf \Gamma}$ (${\bf k}$=(0,0))
to  ${\bf K}$ (${\bf k}=(\frac{4\pi}{3},0)$) and to ${\bf X}$ (${\bf k}=
(\pi,\frac{\sqrt{3}\pi}{3})$) at the critical magnetic fields
$h_{c1}=5.07$rad/ps,  $h_{c2}=6.75$rad/ps, and $h_{c3}=20.39$rad/ps,
respectively. (d)-(f) show the contour maps of the dispersion relation
of band 2 at the three critical magnetic fields. (g)-(i) show the
contour maps of the dispersion relation of band 3 at the three
critical magnetic fields. The squares with number 1, 2, and 3 are marked for the
touching points. In (d), (g) and (e), (h), we only mark one of the
six symmetric points by squares of number 1 and 2 for simplicity.  }
\end{figure}
Figure~\ref{fig5} shows the curves of the derivative of
the phonon Hall conductivity and the Chern numbers at the
critical magnetic fields.
The first derivative of phonon Hall conductivity has a
minimum or maximum at the magnetic fields $h_{c1}=5.07, h_{c2}=6.75,
{\rm and} h_{c3}=20.39$ rad/ps for the finite-size sample (the sample has
$N=N_L^2$ unit cells).
The first derivative $d\kappa_{xy}/dh$ at the points $h_{c1},
h_{c2}, h_{c3}$ diverges when the system size increases to infinity
\cite{zhang10}. At the three critical points the second derivative
$d^2\kappa_{xy}/dh^2$ is discontinuous, which is shown in the inset of
Fig.~\ref{fig5}, across which phase transitions occur.
For different temperatures, the phase transitions occur at exactly the
same critical values. Thus the temperature-independent phase transition
does not come from the thermodynamic effect, but is induced by the topology
of the phonon band structure, which corresponds to the sudden change
of the Chern numbers. While there
is one discontinuity of the Chern numbers for the honeycomb lattice
system, for the kagome lattice system, there are three ones corresponding
to the divergency of the derivative of the phonon conductivity, which can
be seen in Fig.~\ref{fig5}.

The touching and splitting of the phonon bands near the
critical magnetic field induces the abrupt change of Chern
numbers of the phonon band \cite{zhang10}. In Ref.~\cite{zhang10}, for the PHE
in the honeycomb lattices, we know
that band $2$ and $3$ are going to touch with each other at the ${\bf \Gamma}$
point if the magnetic field increases to $h_c$; at the critical
magnetic field, the degeneracy occurs and the two bands possess the
cone shape; above the critical point $h_c$, the two bands split up.
Therefore, the difference between the two bands decreases below and
increases above the critical magnetic field,
and is zero at the critical point. The eigenfrequecy difference is in the
denominator of the Berry curvature,  thus the variation of the difference around
the critical magnetic field, dramatically affects the Berry curvature of the
corresponding bands. In the kagome lattice systems,
we find that the touching and splitting of the phonon bands not only occurs at
the ${\bf \Gamma}$ point, but also occurs at other points, which is shown in
Fig.~\ref{fig6}. At the first
critical points $h_{c1}$, the bands 2 and 3 touch at the point ${\bf K}$
(marked by a square with number 1); at $h_{c2}$ the two bands touch at ${\bf X}$
( marked by a square with number 2); while only for the third
critical one $h_{c3}$, band 2 and 3 degenerate at the point ${\bf \Gamma}$
 (marked by a square with number 3).
From the contour maps of bands 2 and 3, we clearly see that the critical
magnetic fields $h_{c1}$, $h_{c2}$, and $h_{c3}$, there are local maximum
for the band 2 and the local minimum for the band 3. Thus for all
the critical magnetic fields where the Chern numbers have abrupt changes, in the wave-vector space we can always find the phonon bands touching
 and splitting at some symmetric center points.

Therefore, through the study of the PHE in both honeycomb lattices \cite{zhang10} and kagome lattices, we find  discontinuous jumps in Chern numbers, which manifest themselves as singularities of the first derivative of the phonon Hall conductivity with respect to the magnetic field.  Such associated phase transition is connected with the crossing of band 2 and band 3, which corresponds to the touching between a acoustic band and a optical band. However, we can not observe the similar associated phase transition in triangular lattices because of no optical bands, where the Chern number of each band keeps zero while the phonon Hall conductivity is nonzero because of the nonzero Berry curvatures.

\section{Conclusion}
\label{seccon}
We present a new systematic theory of the PHE in the ballistic
crystal lattice system, and give an example application of the PHE in
the kagome lattice which is a model structure of the many
real magnetic materials. By the proper second quantization for the Hamiltonian,
we obtain the formula for the heat current
density, which considers all the phonon bands including both positive and negative frequencies. The heat current density can be divided into two parts,
one is the diagonal, another is off-diagonal. The diagonal part corresponds to the normal velocity; and the  off-diagonal part corresponds to the anomalous velocity which is induced by the Berry vector potential. Such anomalous velocity induces the PHE in the crystal lattice. Based on such heat current density we derive the formula of the phonon Hall conductivity which is in terms of the Berry curvatures. From the application on the kagome lattices, we find that at weak magnetic field, the phonon Hall conductivity changes sign with varying temperatures. It is also found that the mechanism on the PHE about the relation between the phonon Hall conductivity, Chern numbers and the phonon band structure can be generally appllied for kagome lattices.  While there is only one discontinuity in PHE of the honeycomb lattices, in the kagome lattices there are three singularities induced by the abrupt change of the phonon band topology,  which correspond to the touching and
 splitting at three different symmetric center points in the wave-vector space.

\section*{Acknowledgements}
L.Z. thanks Bijay Kumar Agarwalla for fruitful
discussions. J.R. acknowledges the helpful communication
with Hosho Katsura. This project is supported in part by Grants
No. R-144-000-257-112 and No. R-144-000-222-646 of NUS.


\section*{References}
\begin{thebibliography}{01}
\bibitem{Wang2008} Wang L and Li B, Physics World {\bf 21}, No.{\bf 3}, 27
(2008).
\bibitem{wangjs2008} Wang J-S, Wang J, and L\"{u} J T, Eur. Phys. J. B {\bf 62},
381 (2008).
\bibitem{diode} Li B, Wang L, and Casati G, Phys. Rev. Lett. {\bf 93} 184301
(2004); Chang C W, Okawa D, Majumdar A, and Zettl A, Science
{\bf 314}, 1121 (2006).
\bibitem{transistor} Li B, Wang L and Casati G, Appl. Phys. Lett. {\bf 88}, 143501
(2006).
\bibitem{logic} Wang L and Li B, Phys. Rev. Lett. {\bf 99}, 177208 (2007).
\bibitem{memory} Wang L and Li B, Phys. Rev. Lett. {\bf 101}, 267203 (2008).

\bibitem{strohm05} Strohm C, Rikken G L J A, and Wyder P,
Phys. Rev. Lett. \textbf{95}, 155901 (2005).
\bibitem{inyushkin07} Inyushkin A V and Taldenkov A N,
JETP Lett. \textbf{86}, 379 (2007).

\bibitem{shengl06} Sheng L, Sheng D N, and Ting C S,
Phys. Rev. Lett. \textbf{96}, 155901 (2006).
\bibitem{kagan08} Kagan Y and Maksimov L A,
Phys. Rev. Lett. \textbf{100}, 145902 (2008).
\bibitem{wang09} Wang J-S and Zhang L, Phys. Rev. B \textbf{80}, 012301 (2009).
\bibitem{zhang09} Zhang L, Wang J-S, and Li B, New J. Phys. \textbf{11},
113038 (2009).
\bibitem{berry84} Berry M V, Proc. R. Soc. Lond. A \textbf{392}, 45 (1984).
\bibitem{xiao10} Xiao D, Chang M-C, and Niu Q, Rev. Mod. Phys.
 \textbf{82}, 1959 (2010).
\bibitem{TKNN}  Thouless D J, Kohmoto M, Nightingale M P, and den Nijs M,
Phys. Rev. Lett. \textbf{49}, 405 (1982).
\bibitem{kohmoto85} Kohmoto M,  Ann. Phys. \textbf{160}, 343 (1985).
\bibitem{AHE} Nagaosa N, Sinova J, Onoda S, MacDonald A H, and
Ong N P, Rev. Mod. Phys. \textbf{82}, 1539 (2010).
\bibitem{fang03}  Fang Z, Nagaosa N, Takahashi K S, Asamitsu A, Mathieu R, Ogasawara T, Yamada H, Kawasaki M, Tokura Y,
and Terakura K,  Science \textbf{302}, 92 (2003).
\bibitem{xiao06} Xiao D, Yao Y, Fang Z, and Niu Q, Phys. Rev. Lett. \textbf{97}, 026603 (2006).
\bibitem{shengd06} Sheng D N, Weng Z Y, Sheng L, and Haldane F D M,  Phys. Rev. Lett. \textbf{97}, 036808 (2006).
\bibitem{koenig08} Koenig M, Buhmann H, Molenkamp L W, Hughes T, Liu C-X, Qi X-L, and Zhang S-C,  J. Phys. Soc. Jpn. \textbf{77}, 031007 (2008).

\bibitem{prodan09} Prodan E and Prodan C, Phys. Rev. Lett.
{\bf103}, 248101 (2009).
\bibitem{berg11} Berg N, Joel K, Koolyk M, and Prodan E, Phys. Rev. E {\bf 83}, 021913 (2011).
\bibitem{ren10} Ren J, H\"{a}nggi P, and Li B,
Phys. Rev. Lett. {\bf104}, 170601 (2010).
\bibitem{lv10} L\"{u} J-T, Brandbyge M,  and Hedeg{\aa}rd P,
Nano Lett. {\bf10}, 1657 (2010).

\bibitem{zhang10} Zhang L, Ren J, Wang J-S, and Li B,
Phys. Rev. Lett. \textbf{105}, 225901 (2010).
\bibitem{katsura10} Katsura H, Nagaosa N, and Lee P A,
Phys. Rev. Lett. \textbf{104}, 066403 (2010);
Onose Y, Ideue T, Katsura H, Shiomi Y, Nagaosa N, Tokura Y,
Science {\bf 329}, 297 (2010).
\bibitem{holz72} Holz A, Il Nuovo Cimento B \textbf{9}, 83 (1972).
\bibitem{mekata03} Mekata M, Physics Today {\bf 56}, 12(2003).
\bibitem{kagome} Syozi I, Prog. Theor. Phys. {\bf 6}, 306 (1951);
Takano M, Shinjo T, Kiyama M, Takada T, J. Phys. Soc. Jpn.
{\bf 25}, 902 (1968); Wolf M, Schotte K D, J. Phys. A {\bf 21}, 2195 (1988);
 Elser V, Phys. Rev. Lett. 62, {\bf 2405} (1989);
  Broholm C, Appli G, Espinosa G P, Cooper A S,
  Phys. Rev. Lett. {\bf 65}, 3173 (1990).
\bibitem{agarwalla11} Agarwalla B K,  Zhang L, Wang J-S, and Li B,  Eur. Phys. J. B \textbf{81}, 197 (2011).
\bibitem{kronig39} Kronig R de L, Physica (Amsterdam) \textbf{6}, 33 (1939).

\bibitem{vleck40} Van Vleck J H, Phys. Rev. \textbf{57}, 426 (1940).

\bibitem{orbach61} Orbach R, Proc. R. Soc. A \textbf{264}, 458 (1961).

\bibitem{manenkov66} \textsl{Spin-Lattice Relaxation in Ionic Solids},
edited by Manenkov A A and Orbach R (Harper \& Row, New York,
1966).

\bibitem{capellmann87} Capellmann H and Neumann K U, Z. Phys. B {\bf 67}, 53 (1987).

\bibitem{capellmann89} Capellmann H, Lipinski S, and Neumann K U, Z. Phys. B {\bf 75}, 323 (1989).

\bibitem{capellmann91} Capellmann H and Lipinski S, Z. Phys. B {\bf 83}, 199 (1991).

\bibitem{ioselevich95} Ioselevich A S and Capellmann H, Phys. Rev. B {\bf 51},11 446 (1995).

\bibitem{hardy63} Hardy R J, Phys. Rev. \textbf{132}, 168 (1963).
\bibitem{mahan00} Mahan G D, \textsl{Many-Particle Physics} 3rd ed.
(Kluwer Academic, New York, 2000).


\end{thebibliography}
\end{document}